\pdfoutput=1
\documentclass[letterpaper,
               ]{jacow}
\usepackage{pdfpages,multirow,ragged2e, xurl} %

\makeatletter%
	\ifboolexpr{bool{xetex}}
	 {\renewcommand{\Gin@extensions}{.pdf,%
	                    .png,.jpg,.bmp,.pict,.tif,.psd,.mac,.sga,.tga,.gif,%
	                    .eps,.ps,%
	                    }}{}
\makeatother

\ifboolexpr{bool{xetex} or bool{luatex}} 
 {}                                      
 {\usepackage[utf8]{inputenc}}           

\usepackage[USenglish]{babel}

\usepackage{biblatex}

\usepackage{tablefootnote}
\usepackage{fancyhdr}

\begin{document}

\pagestyle{fancy}
\fancyhf{} 
\renewcommand{\headrule}{\smash{\resizebox{0\linewidth}{0.4ex}{\textcolor{cyan}{%
    \fontsize{20}{24}\usefont{U}{webo}{xl}{n}{4}}}}}
\fancyfoot[R]{\vspace*{.2cm}\thepage} 
\DeclareCiteCommand{\supercite}[\mkbibsuperscript]
  {\iffieldundef{prenote}
     {}
     {\BibliographyWarning{Ignoring prenote argument}}%
   \iffieldundef{postnote}
     {}
     {\BibliographyWarning{Ignoring postnote argument}}}
  {\usebibmacro{citeindex}%
   \bibopenbracket\usebibmacro{cite}\bibclosebracket}
  {\supercitedelim}
  {}

\title{Exploring and Applying Audio-Based\\Sentiment Analysis in Music}

\author{Etash Jhanji\thanks{etashjhanji@gmail.com\\Fox Chapel Area High School, Pittsburgh PA USA}}
	
\maketitle
\captionsetup{width=.9\linewidth}
\begin{abstract}
	\textit{
		Sentiment analysis is a continuously explored area of text processing that deals with the computational analysis of opinions, sentiments, and subjectivity of text. However, this idea is not limited to text and speech, in fact, it could be applied to other modalities. In reality, humans do not express themselves in text as deeply as they do in music. The ability of a computational model to interpret musical emotions is largely unexplored and could have implications and uses in therapy and musical queuing. In this paper, two individual tasks are addressed. This study seeks to (1) predict the emotion of a musical clip over time and (2) determine the next emotion value after the music in a time series to ensure seamless transitions. Utilizing data from the Emotions in Music Database, which contains clips of songs selected from the Free Music Archive annotated with levels of valence and arousal as reported on Russel's circumplex model of affect by multiple volunteers, models are trained for both tasks. Overall, the performance of these models reflected that they were able to perform the tasks they were designed for effectively and accurately.}
\end{abstract}

\section{Introduction}
Although sentiment analysis is well researched, studies focus on text-based emotion and represent emotions categorically.\supercite{sentimentSurvey} People often express strong emotions through speech and audio, with more emphasis than in text. Other works focus largely on speech sentiment\supercite{sentimentOnSpeaker}, not necessarily incorporating other sounds in an audio file. A computational model to interpret musical emotions can have medical uses, including therapeutic purposes, and also commercial uses in queueing music for streaming services. This project seeks to explore the capability of Long Short-Term Memory (LSTM) models to (1) predict the emotion of a musical clip over time and (2) determine the next "emotion value" after in a time series to ensure seamless transitions between audio sentiments. 

\section{Background}

\subsection{Representing Emotion}
\begin{figure}[!htb]
	\centering
	\includegraphics*[width=0.8\columnwidth]{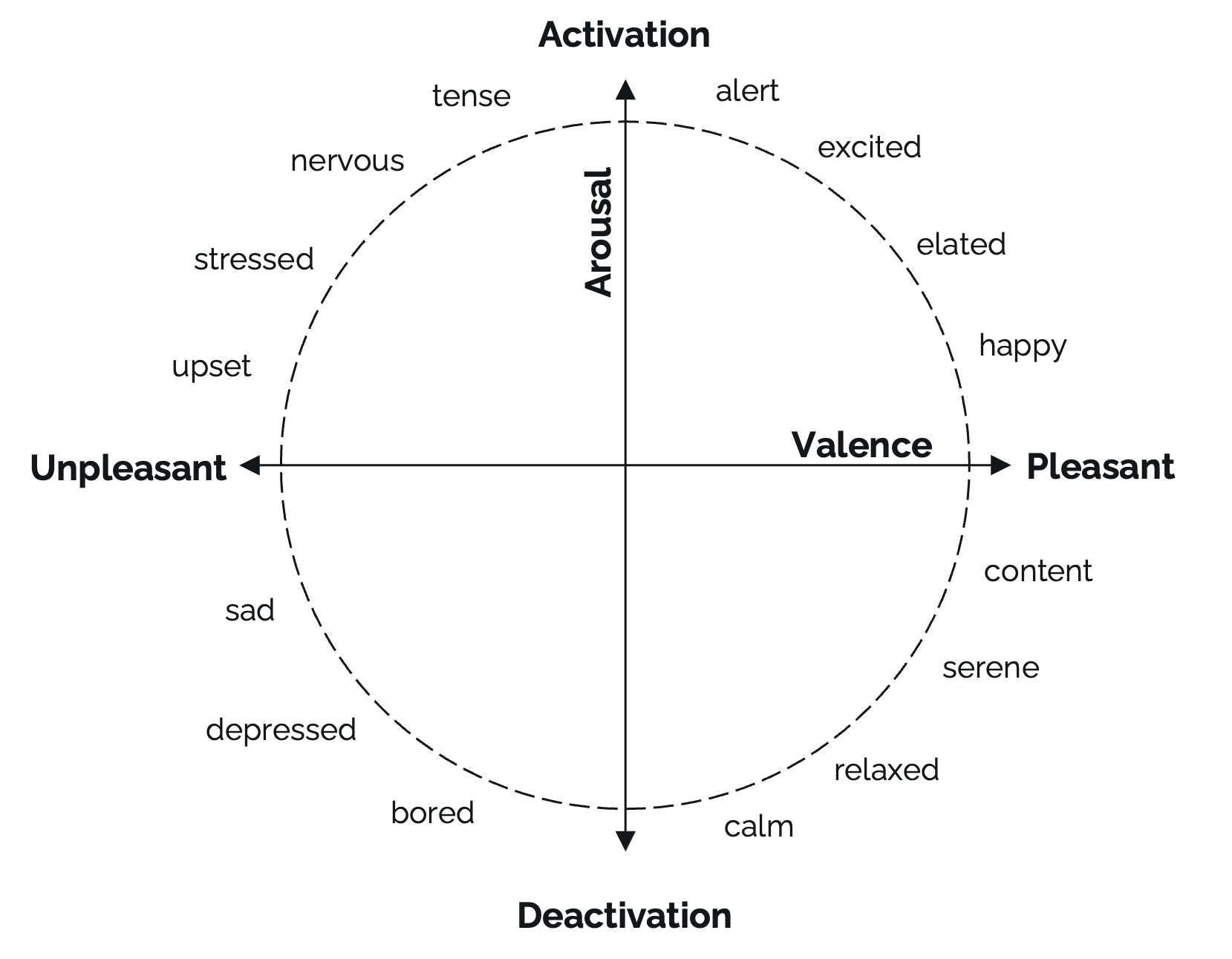}
	\caption{Russel's Circumplex Model of Affect}
	\label{fig:russels_model}
 \end{figure}

Typically, text-based sentiment analysis attempts to classify text into categories of basic emotions as is supported by neuroscience studies on distinct neural circuits for emotions.\supercite{basicEmotion} On the other hand, Russel's circumplex model of affect proposes that emotions arise as a product of two separate circuits: arousal and valence. These can also be renamed as activation and pleasantness, respectively.\supercite{russel} This redefines the sentiment analysis task as a regression task with two outputs rather than a classification task. The database used also utilizes Russel's model on audio samples as the samples are annotated to show the emotion intended to be induced by the clip. Furthermore, the model can be divided into regions to represent emotions with more precision and can additionally describe intensity/severity as can be seen in Fig.~\ref{fig:russels_model}. This is useful in the applications of audio analysis and provides a quantitative measure of emotion. 

\subsection{Audio Processing}
\begin{figure}[htb!]
	\centering
	\includegraphics*[width=1\columnwidth]{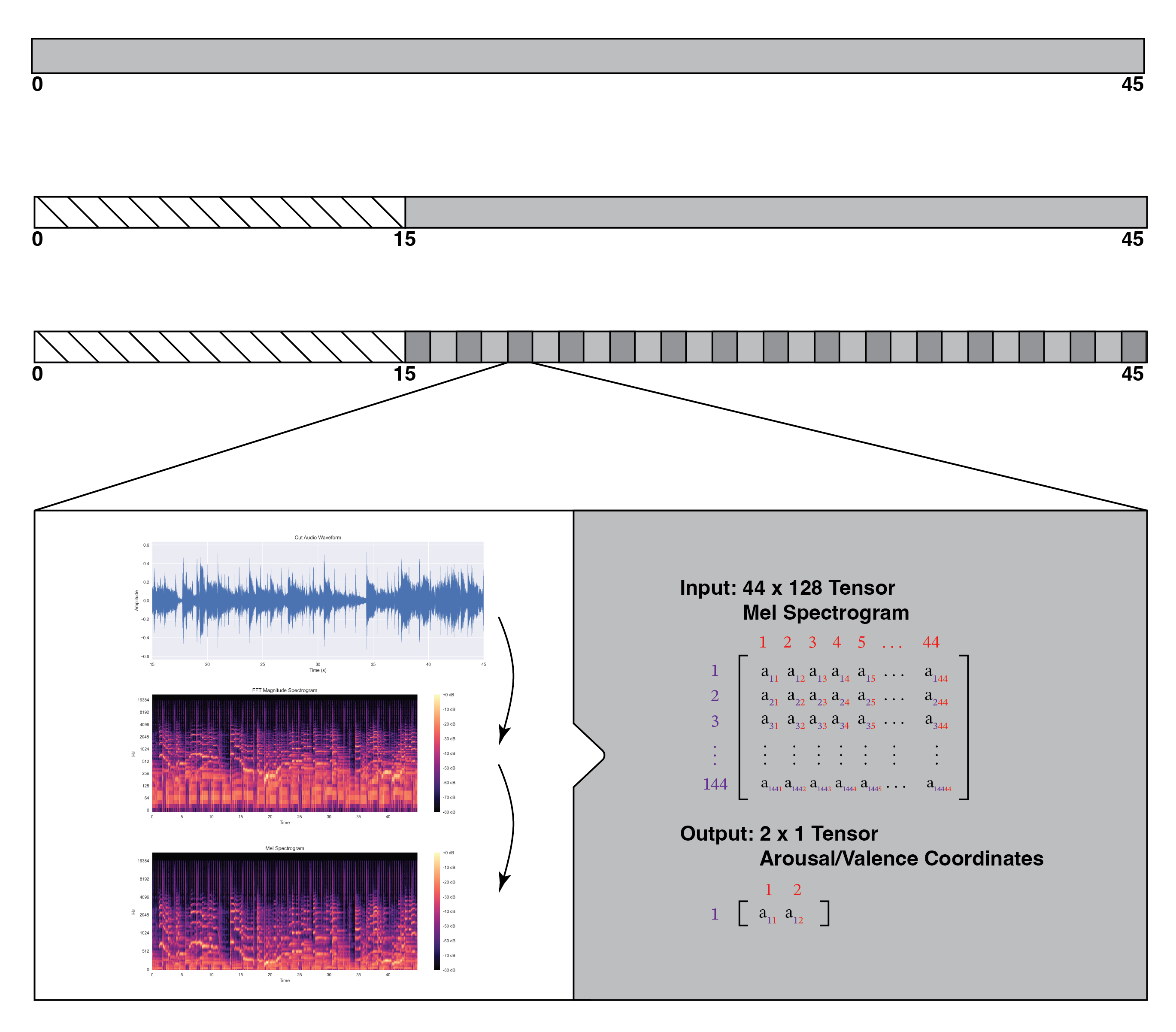}
	\caption{The full pipeline of audio processing including clipping, mel spectrogram, and storage format. }
	\label{fig:full_audio_pipeline}
 \end{figure}

There are many ways to represent audio files for input to a model, but this study uses Logarithmic Mel Spectrograms. This works by taking the raw audio waveform, applying a Fourier Transform, creating a spectrogram of the frequencies, applying the Mel scale to represent human frequencies better, and applying a logarithmic transformation.\supercite{audioProc} Random Gaussian noise is also added to the Mel Spectrogram to improve the robustness of the model. The audio clips were spliced into half-second clips, sampled at 44100 Hz, and the Mel spectrogram has 128 frequency bins, an FFT length of 512, and a hop length of 2048 with Librosa. The output was a NumPy array or PyTorch tensor with a shape of 128 by 44 for 128 bins and 44 timesteps. Alternatively, the model could have used Mel Frequency Cepstral Coefficients (MFCCs) which tend to represent speaker data well\supercite{mfcc}, however, the model should not have to depend on speech features and should instead be able to understand purely instrumentals tracks as well. The extracted Mel Spectrogram dataset was then tied to its continuous arousal and valence coordinates. The full pipeline is demonstrated in Fig.~\ref{fig:full_audio_pipeline}. 

\subsection{Long Short-Term Memory Models}
LSTM models are a subset of recurrent neural networks (RNNs). These models have a hidden state encoding previous information to keep a stateful memory of previous entries to the model. Traditional RNNs have a vanishing gradient problem where previous inputs are forgotten over time, but LSTMs solve this problem with a more sophisticated architecture. This type of model is able to outperform traditional models where long-term dependencies play a part in prediction, such as in audio where data takes place over time.

\subsection{Dataset \supercite{openSMILE, 1000SongforEmotioninMusic}}
The Emotion in Music Database (EmoMusic/1000 songs) consists of about 700 free-use songs that have been annotated by volunteers. The crowdsourced data consists of annotations sampled at 2 Hz (every 0.5 seconds) for arousal and valence of a clip on a scale from -1 to 1 each. The continuously sampled data points are the average of 10 volunteers' annotations at that instant to ensure stability of data. The standard deviations of the continuous annotations are also provided, averaging about 0.35 across all samples. This averaged standard deviation is the goal for the first model's root mean square error (RMSE) because it means the model's variation can be accounted for by the human variation in the dataset. The first 15 seconds of annotations in the data were dropped due to instability and the 45-second clips of audio are provided with the dataset. 

\section{Task 1: Predicting Emotion}

\subsection*{Input/Output}
The first task is to predict the arousal and valence values of a 0.5-second clip of audio's mel spectrogram. The input is a tensor of shape 128 by 44 representing the clips' mel spectrogram with specific settings (described in the background). The output will be a two-value tensor representing arousal and valence. 

\subsection*{Evaluation Metric}
The model will use mean squared error (MSE) for its evaluation metric and loss function. MSE penalizes heavily for deviation as it the square of the output unit. The target for model loss is 0.09 (or RMSE of 0.3) as described in the background section. Since human annotations vary by about 0.3 on average on the same songs, the error produced by the model can be accounted for by human error in the dataset. 

\subsection*{Hyperparameter Optimization}
\begin{figure}[!ht]
	\centering
	\includegraphics*[width=0.85\columnwidth]{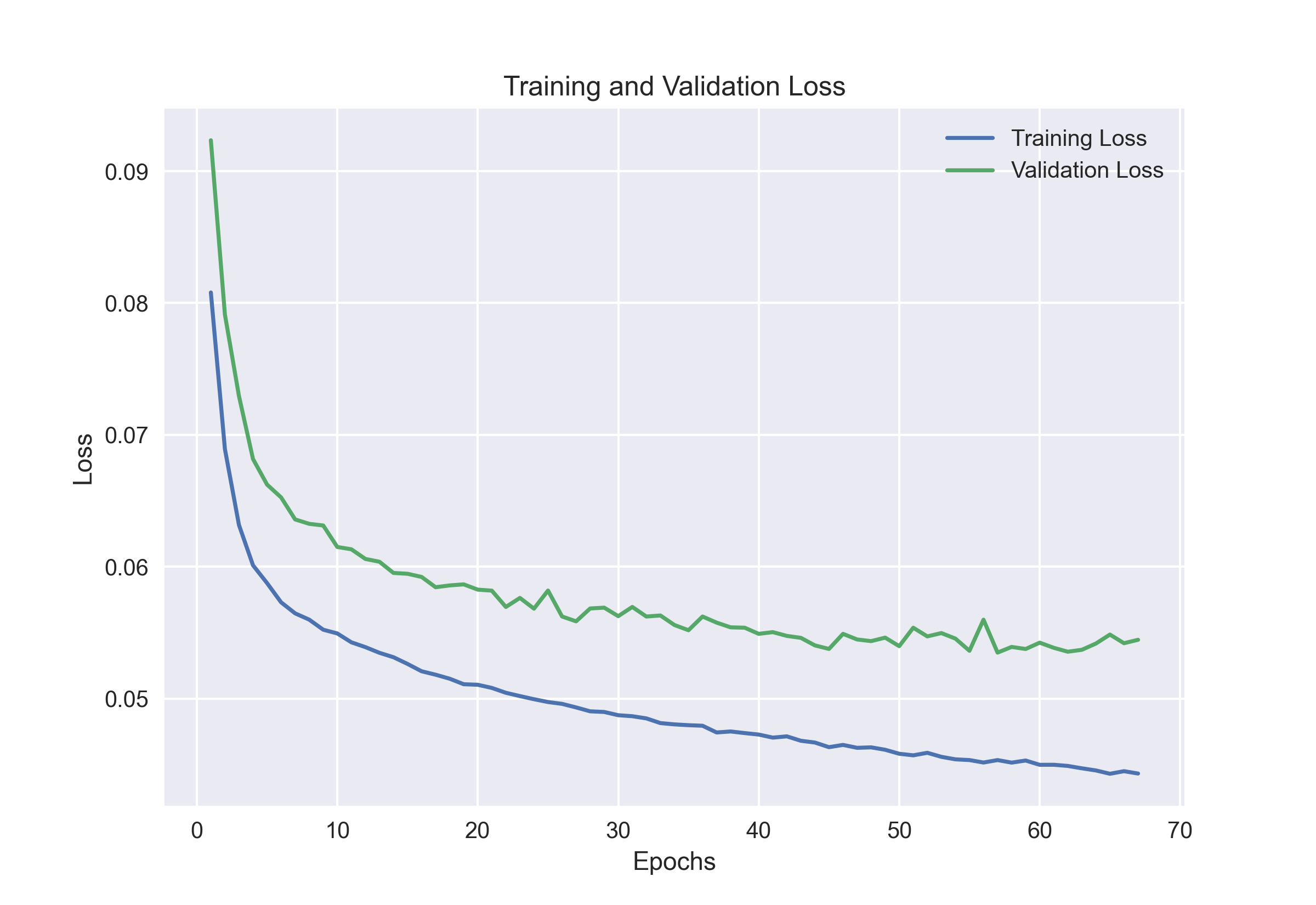}
	\caption{Loss graphs (MSE) for training and validation varied with number of epochs for task 1 using the most optimal hyperparameter found. Shows convergence for training but slight possible overfitting. }
	\label{fig:task1loss}
 \end{figure}

This model uses the Adam optimizer. Experimenting with hyperparameters, the learning rate had the highest influence on results, and a low number of hidden layers performed best. To determine the number of epochs an early stopping technique was used in which the model automatically ended training after loss converged and stopped decreasing. A fixed batch size of 58 samples was used, where each batch contained every clip from one audio file. The best results found are shown in Table.~\ref{tab:task1results} and Fig. \ref{fig:task1loss}. Although decreasing the learning rate from \num{5e-5} may slightly improve the convergence of MSE, the final MSE is very similar whereas training time and epochs increase. 

\begin{table}[!hbt]
	\centering
	\caption{Task 1 Best Results}
	\begin{tabular}{lcc}
		\toprule
		\textbf{Parameter} & \textbf{Value} \\
		\midrule
			Learning Rate         & \num{5e-5} \\
		    Hidden Size           & 20 \\
			No. of Modules\tablefootnote{Where a "module" is an nn.LSTM object in the model and layers within a module are stacked}     & 2 \\
			No. of Layers (per module)         & 2 \\
			No. of Layers (total)       & 4 \\
			Dropout Probability   & 0.1 \\
			Epochs                & 67 \\
		\bottomrule
			Train MSE                   & 0.044 \\
			Validation MSE                  & 0.054 \\
	\end{tabular}
	\label{tab:task1results}
 \end{table}
 \vspace*{-20pt}

 \section{Task 2: Intelligent Queuing}
 \subsection*{Input/Output}
 The second task is to predict the next pair of arousal and valence values from an arbitrarily chosen time sequence of length 10. The input is a tensor of shape 2 by 10 representing the arousal and valence over ten timesteps. The output will be a two-value tensor representing the next (11th) arousal and valence point. 
 
 \subsection*{Evaluation Metric}
 The model will also use MSE for its evaluation metric and loss function for the same reasons as in Task 1. The target MSE/RMSE are similar due to natural variation in humans but are expected to be lower in a simpler task. 
 
 \subsection*{Hyperparameter Optimization}
\begin{figure}[h!]
	\centering
	\includegraphics*[width=0.85\columnwidth]{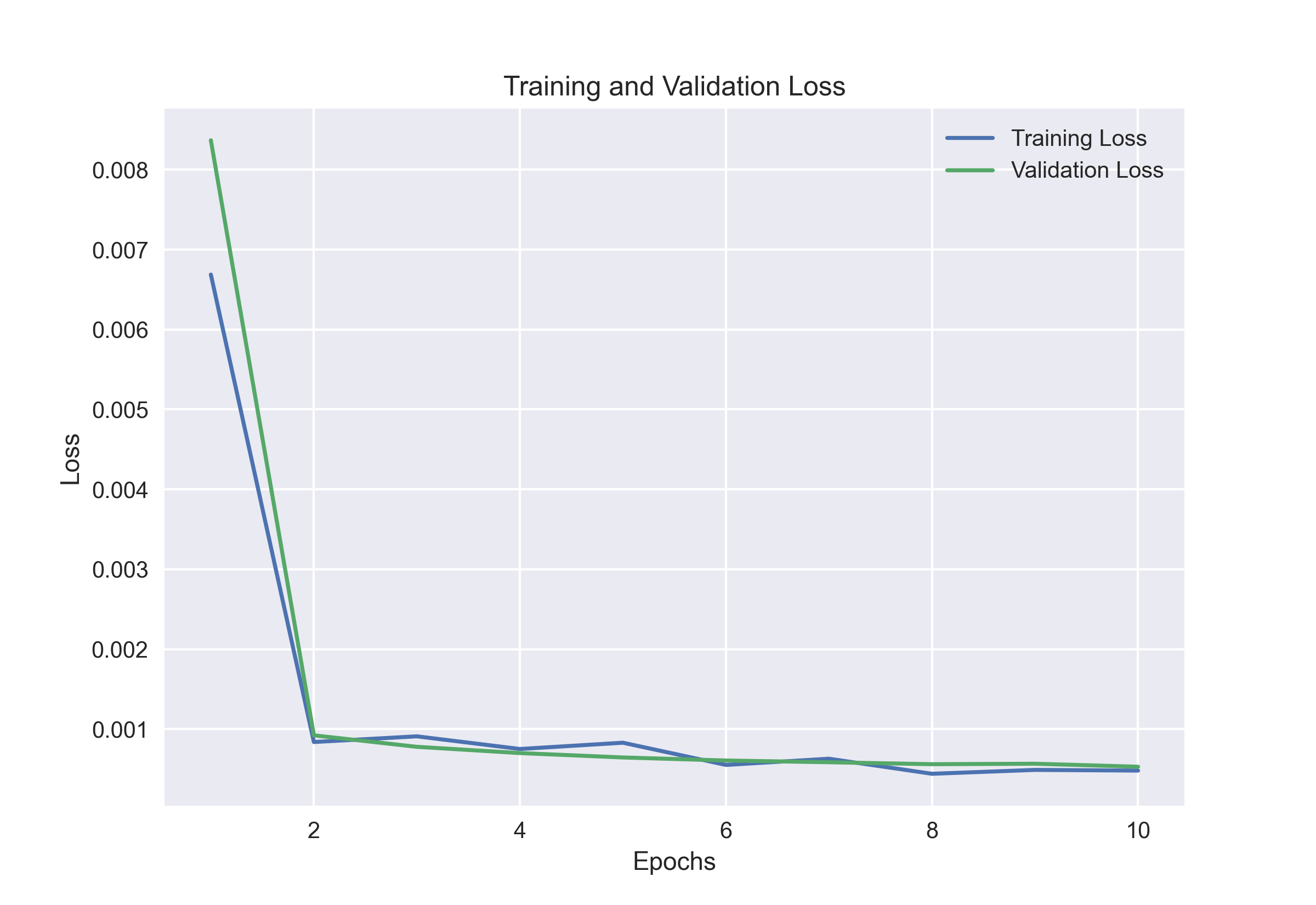}
	\caption{Loss graphs (MSE) for training and validation varied with number of epochs for task 2 using the most optimal hyperparameter found. The loss converges very quickly after the first two epochs.}
	\label{fig:task2loss}
 \end{figure}

 Similar to the previous task, hyperparameter optimization was largely experimental with learning rate and epoch count having a high influence on results. This model still uses the Adam optimizer and mean squared error, however, batching is randomized and the task in simpler. The best results found are shown in Table.~\ref{tab:task2results} and Fig.~\ref{fig:task2loss}.

 \begin{table}[!hbt]
	 \centering
	 \caption{Task 2 Best Results}
	 \begin{tabular}{lcc}
		 \toprule
		 \textbf{Parameter} & \textbf{Value} \\
		 \midrule
			 Learning Rate         & \num{1e-4} \\
			 Hidden Size           & 32 \\
			 Batch Size           & 64 \\
			 No. of Modules\tablefootnote{Where a "module" is an nn.LSTM object in the model and layers within a module are stacked}     & 1 \\
			 No. of Layers (per module)         & 2 \\
			 No. of Layers (total)       & 42\\
			 Dropout Probability   & 0 \\
			 Epochs                & 10 \\
		 \bottomrule
			 Train MSE                   & \num{4e-4} \\
			 Validation MSE                  & \num{5e-4} \\
	 \end{tabular}
	 \label{tab:task2results}
  \end{table}
  \vspace*{-20pt}

\subsection{Linear Regression Approach}
\begin{figure}[!htb]
	\centering
	\includegraphics*[width=0.65\columnwidth]{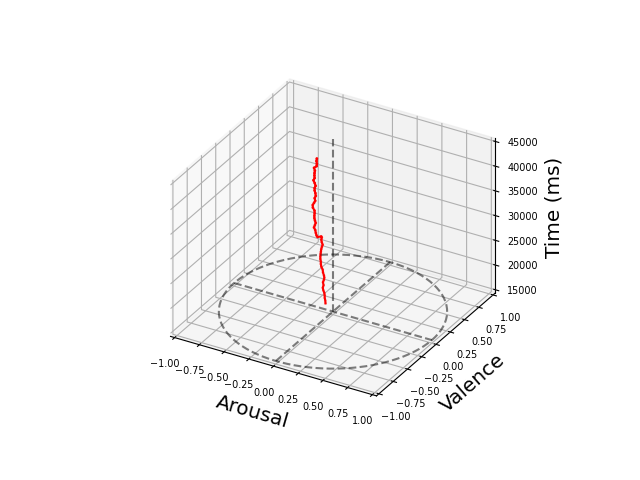}
	\caption{Arousal and valence for one song plotted against time on the z-axis. Dotted lines represent the centerlines and edges of the circumplex model. }
	\label{fig:3drusselplot}
 \end{figure}

\begin{figure}[!htb]
	\centering
	\includegraphics*[width=0.65\columnwidth]{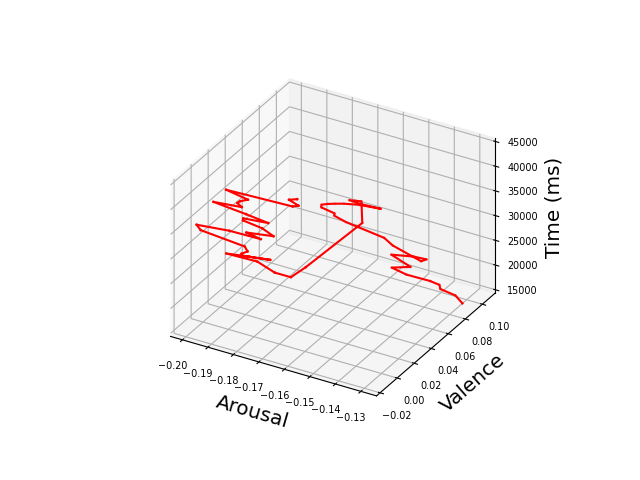}
	\caption{A zoomed view of Figure \ref{fig:3drusselplot} that shows more erratic and nonlinear patterns in arousal and valence. }
	\label{fig:3drusselplotzoomed}
 \end{figure}

\begin{figure}[!htb]
	\centering
	\includegraphics*[width=0.95\columnwidth]{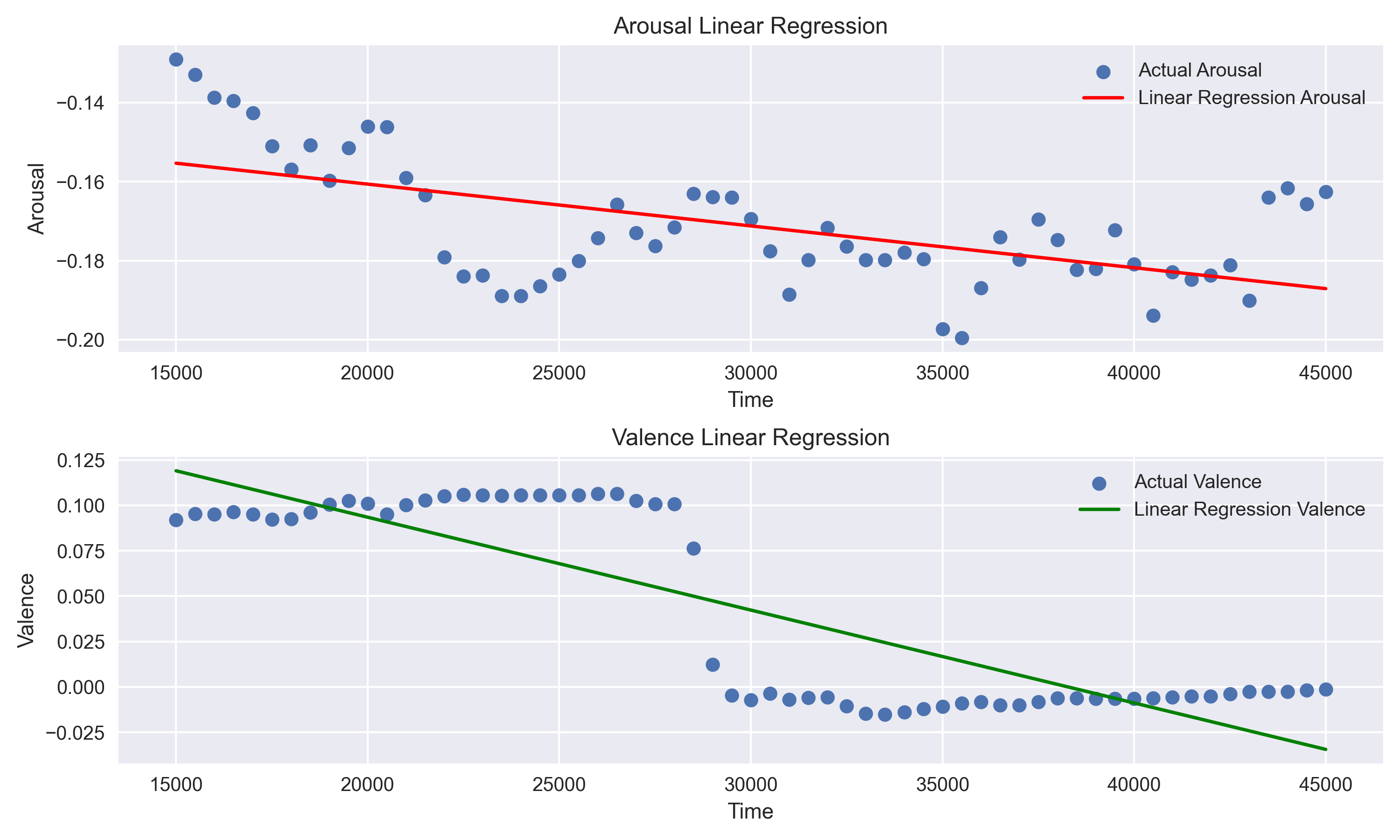}
	\caption{The linear regression results of one particular song. Shows a good model of the trend but a clear lack of accuracy around the center of the time axis. }
	\label{fig:linreg}
 \end{figure}

Since the LSTM results showed that the task may have been too simple and revealed a possible exploding gradient issue, instead a linear regression approach was used. One in which arousal was the dependent variable and the other in which valence was the dependent variable with both having time as the independent variable. When the arousal and valence are plotted against time, they appear linear/constant (Fig. \ref{fig:3drusselplot}), but, zooming in, some clips show erratic, nonlinear behavior (Fig. \ref{fig:3drusselplotzoomed}). 
The linear regression showed that the data's overall trend may be able to be modeled by a regression, however, the model was not suitable for predicting exact values, especially within the songs (Fig. \ref{fig:linreg}). 

\vspace*{-15pt}
 \vspace*{-20pt}

 \vspace*{-10pt}

 \section{Results and Conclusions}
 \subsection{Task 1}
 The emotion prediction model got an MSE loss of about 0.055 in validation and 0.044 in training meaning that RMSE are 0.235 and 0.21, respectively. These RMSE values can be accounted for by the natural variation in even human choice annotation of this data, indicating the model is doing well, matching human predictions. 
 \subsection{Task 2}
 The "next value" predictor model showed an MSE of about 0.0004 in training and 0.0005 in validation. These results indicate that the model performed well. The linear regression model proved to be less effective in predicting exact values but may be effective in considering general trends. 

\section{Discussion}
The ability to automatically determine quantitative values to describe emotion of a music clip can have uses in the real world. In the medical field, music is well known as having psychological effects on mood but has also been found to have effects on other neurological disorders including multiple sclerosis and Parkinson's. \supercite{therapy} Commercial applications of this software also include intelligent queueing of music to maintain a smooth flow of mood in music playback. A demonstrative implementation of this model is included in the public repository.\footnote{Please reference the GitHub repository for more information in the \textit{demo.py} file} 

\section{Future Direction}
For task 1 a broad, automated hyperparameter search may be able to refine and optimize hyperparameter for even better regression results. Different batching strategies and model complexities may also result in an overall superior model. The second task would benefit from many of the improvements of task 1, as well as using a statistic to objectively measure the performance of different models such as the linear regression. Adding variable length time-sequence input is also a possibility. Implementing this code in an open-source library that can be used in applications could greatly help those who use it. The code could also be implemented alongside the queuing algorithms of major streaming services (e.g. Spotify) to offer improved experiences to users.

\section{Error Analysis}
Limited hyperparameter scope, input format, noise added, and human error could have resulted in errors in both task one and two's LSTM models. In model one it is important to note the slight overfitting occurring indicating that hyperparameters could likely be further optimized and in model two there may be an exploding gradient problem. In the second model, the predicted value tended to hold the arousal and valence constant due to the mostly constant nature of data, however, in songs with greater changes, the model reflected those changes. In implementation, adding a tolerance to allow for changes between songs would avoid holding the arousal and valence constant unintentionally.

\section{Data Availability}
The Emotion in Music Database (1000 songs) dataset is available online after filling out a request form.\supercite{openSMILE, 1000SongforEmotioninMusic} 

\section{Code Availability}
The code, models, and results from this dataset have been open-sourced under the MIT license and are available at \url{https://github.com/etashj/Exploring-and-Applying-Audio-Based-Sentiment-Analysis}. 

\section{Acknowledgments}
I would like to thank the developers of the Emotions in Music Database\supercite{openSMILE, 1000SongforEmotioninMusic} for making their code open source and free to use as it was of great importance to the development of this project. I would also like to thank the developers of the PyTorch\supercite{pytorch}, NumPy\supercite{numpy}, pandas\supercite{pandas}, and librosa\supercite{librosa} libraries for the Python programming language who also made their code open source which provided the framework for this project. 

%

\printbibliography
%
%


\end{document}